\documentclass[12pt]{article}
\usepackage{latexsym}
\begin{document}
\begin{center}
{\bf A BRIEF REVIEW OF THE SINGULARITIES IN 4D AND 5D VISCOUS
COSMOLOGIES NEAR THE FUTURE SINGULARITY }

\vspace{1cm} I. Brevik\footnote{E-mail: iver.h.brevik@ntnu.no}

\bigskip

Department of Energy and Process Engineering, Norwegian University
of Science and Technology, N-7491 Trondheim, Norway

\bigskip

O. Gorbunova\footnote{E-mail: gorbunovaog@tspu.edu.ru}

\bigskip

Tomsk State Pedagogical University, Tomsk, Russia

\vspace{1cm} %\today
\end{center}

\begin{abstract}

Analytic properties of physical quantities in the cosmic fluid
such as energy density $\rho(t)$ and Hubble parameter $H(t)$ are
investigated near the future singularity (Big Rip). Both 4D and 5D
cosmologies are considered  (the Randall-Sundrum II model in the
5D case), and the fluid is assumed to possess a bulk viscosity
$\zeta$. We consider both Einstein gravity and modified gravity,
where in the latter case the Lagrangian contains a term $R^\alpha$
with $\alpha$ a constant. If $\zeta$ is proportional to the power
$(2\alpha-1)$ of the scalar expansion, the fluid can pass from the
quintessence region into the phantom region as a consequence of
the viscosity. A property worth noticing is that the 4D
singularity on the brane becomes carried over to the bulk region.

\end{abstract}

\section{Introduction}

The possibility of crossing the $w=-1$ barrier in dark energy
cosmology has recently become a topic of considerable interest.
One usually assumes that the equation of state for the cosmic
fluid can be written in the form
\begin{equation}
p=w\rho, \label{1}
\end{equation}
where $w$  is a constant. If $w=-1$ the fluid is called a "vacuum
fluid", with peculiar thermodynamic properties such as negative
entropy \cite{brevik04}. More general forms for the equation of
state can be envisaged, such as
\begin{equation}
p=w(\rho)\rho = -\rho-f(\rho), \label{2}
\end{equation}
which is a form that we shall consider below. As is know,
cosmological observations indicate that the present universe is
accelerating. Recent discussions on the actual value of $w$ can be
found, for instance, in refs.~\cite{vikman05,capozziello06,wei08}.
Perhaps, $w$ is even an oscillating function in time. For
discussions on time-dependent values of $w$, one may consult
Refs.~\cite{nojiri06,brevik07,brevik07a}. The possibility of
crossing from the quintessence region ($-1<w<-1/3$) into the
phantom region $w<-1$, is obviously of physical interest. It may
be noted that both quintessence and phantom fluids lead to the
inequality $\rho +3p \leq 0 $, thus breaking the strong energy
condition.

Once being in the phantom region, the cosmic fluid will inevitably
be led into a future singularity, called the Big Rip
\cite{caldwell03,mcinnes02,barrow04,nojiri05}. And this brings us
to the main theme of the present paper, namely to give an overview
of the behavior of central physical quantities near the future
singularity. This  is the case of main  interest. We think that
such an exposition should be useful, not least so because the
situation is rather complex. Namely, there is a variety of
different factors at play here: (i) the thermodynamic parameter
$w(\rho)$, (ii) the possible time dependence of the bulk
viscosity, $\zeta =\zeta(t)$, and (iii) the adoption of Einstein's
gravity, or a version of the so-called modified gravity. (For an
introduction to modified gravity theories, one may consult
Refs.~\cite{nojiri07,cognola08}.)

To begin with, it is convenient to quote from Ref.~\cite{nojiri05}
the classification of possible future singularities:

(i) Type I ("Big Rip"): For $t \rightarrow t_s,~ a\rightarrow
\infty,~\rho \rightarrow \infty$, and $|p|\rightarrow \infty$, or
$p$ and $\rho$ are finite at  $t=t_s$.

(ii) Type II ("sudden"): For $t\rightarrow t_s,~a\rightarrow
a_s,~\rho \rightarrow \rho_s$, and $|p|\rightarrow \infty$,

(iii) Type III: For $t\rightarrow t_s,~a\rightarrow a_s,~\rho
\rightarrow \infty$, and $|p|\rightarrow \infty$,

(iv) Type IV: For $t\rightarrow t_s,~a\rightarrow a_s,~\rho
\rightarrow 0,~|p|\rightarrow 0$, or $p$ and $\rho$ are finite.
Higher order derivatives of $H$ diverge.

\vspace{0.5cm}

Here the notation is standard, $a$ meaning the scale factor and
$t_s$ referring to the instant of the singularity. The above
classification was introduced in the context of ideal, i.e.,
nonviscous, cosmology. We can however make use of the same
classification also in the viscous case.

In the following, we will present salient features of 4D,
respective 5D, viscous cosmology theory, and thereafter focus on
the classification of the various alternatives.

\section{Viscous 4D theory: Basics}

We include this basic material mainly for reference purposes. We
consider the standard FRW metric,
\begin{equation}
ds^2=-dt^2+a^2(t)d{\bf x}^2, \label{3}
\end{equation}
and set the spatial curvature $k$, as well as the 4D cosmological
constant $\Lambda_4$, equal to zero. The Hubble parameter is
$H=\dot{a}/a$, the scalar expansion is $\theta={U^\mu}_{;\mu}=3H$
with $U^\mu$ the four-velocity of the fluid, and $\kappa_4^2=8\pi
G_4$ is the gravitational coupling. Of main interest are the
$(tt)$ and $(rr)$ components of the Friedmann equations. They are
\begin{equation}
\theta^2=3\kappa_4^2 \,\rho, \label{4}
\end{equation}
\begin{equation}
\frac{2\ddot{a}}{a}+H^2=-\kappa_4^2\,\tilde{p}, \label{5}
\end{equation}
where $\tilde{p}=p-\zeta \theta$ is the effective pressure. From
the differential equation for energy, ${T^{0\nu}}_{;\nu}=0$, we
get
\begin{equation}
\dot{\rho}+(\rho+p)\theta=\zeta \theta^2. \label{6}
\end{equation}
As a consequence of positive entropy change in an irreversible
process, we must require the value of $\zeta$ to be non-negative.
From the equations above we obtain the following  differential
equation for the scalar expansion,
\begin{equation}
\dot{\theta}-\frac{f(\rho)}{2\rho}\theta^2-\frac{3}{2}\kappa_4^2\,\zeta
\theta=0. \label{7}
\end{equation}
In view of the relationship
$\dot{\theta}=(\sqrt{3}\,\kappa_4/2)\dot{\rho}/\sqrt{\rho}$ we can
alternatively reformulate this equation as an equation for the
density,
\begin{equation}
\dot{\rho}-\kappa_4\,\sqrt{3\rho}\,f(\rho)-3\kappa_4^2\,\zeta
\rho=0. \label{8}
\end{equation}
The solution is (cf. Eq.~(9) in \cite{brevik05})
\begin{equation}
t=\frac{1}{\sqrt{3}\,\kappa_4}\int_{\rho_*}^\rho
\frac{d\rho}{\sqrt{\rho}f(\rho)[1+\kappa_4\,\zeta
\sqrt{3\rho}/f(\rho)]}. \label{9}
\end{equation}
We here let $t=0$ be the initial (present) time, and let the
corresponding initial density be $\rho_*$. The functional form of
the bulk viscosity $\zeta$ is so far unspecified. The shear
viscosity is omitted, due to  the assumed spatial isotropy in the
cosmic fluid. Note the dimensions: $ [\kappa_4^2]={\rm
cm}^2,~[f(\rho)]=[\rho]={\rm cm}^{-4}, ~[\zeta]={\rm cm}^{-3}$.
Viscous cosmology are treated at various places, for instance, in
Refs.~\cite{capozziello06,weinberg71,padmanabhan87,gron90,brevik94,hallanger,borkje,borven,ren06,brevik06,sussman08,brevik08c}.

\section{Specific cases in 4D}

We are now in a position to discuss various cases in 4D
explicitly. We have to distinguish between several alternatives:
(i) use of Einstein or modified gravity; (ii)  possible density
dependence of the thermodynamic parameter $w=w(\rho)$; and (iii)
possible time dependence of the bulk viscosity $\zeta(t)$.

\subsection{Einstein gravity, $w$ and $\zeta$ being constants}

Let $f(\rho)=\alpha \rho$, with $\alpha$ a constant. The equation
of state is then
\begin{equation}
p=w\rho=-(1+\alpha)\rho. \label{10}
\end{equation}
We can now solve explicitly for the Hubble parameter
\cite{brevik05,brevik08},
\begin{equation}
H(t)=\frac{H_*e^{t/t_c}}{1-\frac{3\alpha}{2}H_*t_c(e^{t/t_c}-1)},
\label{11}
\end{equation}
where $H_*$ is the present-time value of $H$ and $t_c$ is the
'viscosity time',
\begin{equation}
t_c=\left( \frac{3}{2}\kappa_4^2\zeta\right)^{-1}. \label{12}
\end{equation}
From Eq.~(\ref{11}) it is seen that $H(t)$ becomes singular when
the denominator vanishes. Let us first for reference purposes set
$\zeta=0$:

{\it The nonviscous case}. If $t_{s0}$ designates the singularity
time, we have
\begin{equation}t_{s0}=\frac{2}{3\alpha H_*}. \label{13}
\end{equation}
Then \cite{brevik08},
\begin{equation}
H(t)=\frac{H_* t_{s0}}{t_{s0}-t}, \label{14}
\end{equation}
\begin{equation}
a(t)=\frac{a_* t_{s0}^{2/3\alpha}}{(t_{s0}-t)^{2/3\alpha}},
\label{15}
\end{equation}
\begin{equation}
\rho(t)=\frac{\rho_* t_{s0}^2}{(t_{s0}-t)^2}. \label{16}
\end{equation}

{\it The viscous case}.  If now $t_{s\zeta}$ denotes the
singularity time, we get from Eq.~(\ref{11})
\begin{equation}
t_{s\zeta}=t_c\ln \left[ 1+\frac{2}{3\alpha}\frac{1}{H_*
t_c}\right]. \label{17}
\end{equation}
and
\begin{equation}
H(t)\rightarrow \frac{H_*t_{s0}}{t_{s0}-t}, \quad t\rightarrow
t_{s\zeta}. \label{18}
\end{equation}
Close to the singularity we thus obtain the same singular behavior
as in the nonviscous case. Moreover, we get the following forms,
\begin{equation}
a(t)\sim (t_{s\zeta}-t)^{-2/3\alpha}, \quad t\rightarrow
t_{s\zeta}, \label{19}
\end{equation}
\begin{equation}
\rho(t) \sim (t_{s\zeta} -t)^{-2}, \quad t\rightarrow t_{s\zeta}.
\label{20}
\end{equation}
The viscosity tends to shorten the singularity time,
\begin{equation}
t_{s\zeta}<t_{s0}, \label{21}
\end{equation}
but it does {\it not} modify the exponents in the singularity. The
singularity is of Type I if $\alpha >0$, and of Type II if $\alpha
<0$.

\subsection{Einstein gravity, $f(\rho)=A\rho^\beta$, and $\zeta$ being
constant}

We shall assume that $\beta \geq 1$. From Eq.~(\ref{9}) it is
apparent that the last term in the denominator dominates for large
$\rho$. Near the singularity we obtain  the form
\begin{equation}
\rho(t) \sim (t_{s\zeta}-t)^{\frac{-2}{2\beta-1}}, \quad
t\rightarrow t_{s\zeta}, \label{22}
\end{equation}
which generalizes Eq.~(\ref{20}) and reduces to it when $\beta=1$.
Thus $\rho \rightarrow \infty$ implying, according to
Eq.~(\ref{2}), that also $|p|\rightarrow \infty$. The Hubble
parameter becomes
\begin{equation}
H(t) \sim (t_{s\zeta}-t)^{\frac{-1}{2\beta-1}}. \label{23}
\end{equation}
If $\beta >1$, $a\rightarrow a_s$ (a finite value)  when
$t\rightarrow t_{s\zeta}$. The singularity is of Type III. If
$\beta=1$, the singularity is of Type I.

The material of this subsection was discussed  also in
Ref.~\cite{brevik08}, whereas the equation of state corresponding
to (\ref{22}) and(\ref{23}) was discussed  by Nojiri and Odintsov
\cite{nojiri04}.

\subsection{Modified gravity, $w$ being constant, and
$\zeta=\tau \theta^{2\alpha -1}$}

Consider now the following gravity model,
\begin{equation}
S=\frac{1}{2\kappa_4^2}\int d^4 x \sqrt{-g}(f_0R^\alpha +L_m),
\label{24}
\end{equation}
where $f_0$ and $\alpha$ are constants, $L_m$ being the matter
Lagrangian. This model has been considered before; cf., for
instance, Refs.~\cite{abdalla05,brevik05a,brevik06a,brevik06}. The
case $f_0=1$ and $\alpha=1$ yields Einstein's gravity. The
equations of motion following from the action above are
\begin{equation}
 -\frac{1}{2}f_0g_{\mu\nu}R^\alpha +\alpha
f_0R_{\mu\nu}R^{\alpha-1}-\alpha f_0\nabla_\mu\nabla_\nu
R^{\alpha-1} +\alpha f_0 g_{\mu\nu}\nabla^2
R^{\alpha-1}=\kappa_4^2 T_{\mu\nu}, \label{25}
\end{equation}
where $T_{\mu\nu}$ corresponds to the term $L_m$ in the
Lagrangian. For the cosmic fluid we have
\begin{equation}
T_{\mu\nu}=\rho U_\mu U_\nu+\tilde{p}h_{\mu\nu}, \label{26}
\end{equation}
where $h_{\mu\nu}=g_{\mu\nu}+U_\mu U_\nu$ is the projection tensor
and $\tilde{p}=p-\zeta \theta$ the effective pressure. In comoving
coordinates, $U^0=1,\,U^i=0$. We assume now the simple equation of
state given in Eq.~(\ref{1}).

Of main interest is the (00)-component of Eq.~(\ref{25}). Using
that $R=6(\dot{H}+2H^2)$, $T_{00}=\rho$, as well as the energy
conservation equation (\ref{6}) which in turn follows from
$\nabla^\nu T_{\mu\nu}=0$, we obtain
\[ \frac{3}{2}\gamma f_0 R^\alpha +3\alpha f_0[2\dot{H}-3\gamma
(\dot{H}+H^2)]R^{\alpha-1}+3\alpha (\alpha-1)f_0 [(3\gamma
-1)H\dot{R}+\ddot{R}]R^{\alpha-2} \]
\begin{equation}
+3\alpha (\alpha-1)(\alpha-2)f_0\dot{R}^2 R^{\alpha-3}=9\kappa_4^2
\zeta H, \label{27}
\end{equation}
with $\gamma=w+1$.  The important point now is that this
complicated equation for $H(t)$ is satisfied with the following
form
\begin{equation}
H=H_*/X, \quad {\rm where} \quad X=1-BH_*t, \label{28}
\end{equation}
$B$ being a nondimensional parameter. For Big Rip to occur, $B$
has to be positive.

Taking the bulk viscosity to have the form
\begin{equation}
\zeta=\tau \theta^{2\alpha-1}=\tau (3H)^{2\alpha -1} \label{29}
\end{equation}
with $\tau$ a positive constant, the time-dependent factors in
Eq.~(\ref{27}) drop out. There  remains an algebraic equation,
determining $B$.

Of main interest is the time-dependent forms
\begin{equation}
\zeta=\tau (3H_*/X)^{2\alpha-1}, \quad \rho=\rho_*/X^{2\alpha}.
\label{30}
\end{equation}
As an example, the case $\alpha=2$ turns out to yield a cubic
equation for $B$. There is one positive root (assuming $f_0$
positive), leading to a viscosity-generated Big Rip. If $\alpha
<0$, typically $\alpha=-1$, there may still be positive solutions
for $B$ implying that $H=H_*/X$ is diverging. By contrast, $\zeta
\propto X^{-(2\alpha-1)}$ and $\rho \propto X^{-2\alpha}$ go to
zero.

\section{Relationship to 5D viscous theory}

Let us investigate the possible link between the 4D theory above
and the analogous viscous theory in 5D space. To this end we
consider a spatially flat ($k=0$) brane located at the fifth
dimension $y=0$, surrounded by an anti-de-Sitter (AdS) space. If
the 5D cosmological constant, called $\Lambda$, is negative, the
configuration is that of the Randall-Sundrum II model (RSII)
\cite{randall99}. The 5D coordinates are denoted $x^A=(t, {\bf x},
y)$, and the 5D gravitational coupling is $\kappa_5^2=8\pi G_5$.
The Einstein equations are
\begin{equation}
R_{AB}-\frac{1}{2}g_{AB}R+g_{AB}\Lambda=\kappa_5^2 \,T_{AB},
\label{31}
\end{equation}
and the metric is
\begin{equation}
ds^2=-n^2dt^2+a^2\delta_{ij}dx^idx^j+dy^2, \label{32}
\end{equation}
where $n(t,y)$ and $a(t,y)$ are to be determined from the Einstein
equations.

Of main interest are the $(tt)$ and $(yy)$ components of the field
equations. They are
\begin{equation}
3\left\{ \left( \frac{\dot{a}}{a}\right)^2 -n^2
\left[\frac{a''}{a} +\left(\frac{a'}{a} \right)^2 \right]
\right\}-\Lambda n^2=\kappa_5^2 \,T_{tt}, \label{33}
\end{equation}
\begin{equation}
3\Bigg\{\frac{a'}{a}\left(\frac{a'}{a}+\frac{n'}{n}\right)-\frac{1}{n^2}\Big[\frac{\dot{a}}{a}\left(
\frac{\dot{a}}{a}-\frac{\dot{n}}{n}\right)
+\frac{\ddot{a}}{a}\Big]\Bigg\}+\Lambda=\kappa_5^2 \,T_{yy}
\label{34}
\end{equation}
(cf. for instance,
Refs.~\cite{binetruy00,binetruy00a,brevik02,hallanger,brevik08}).
Overdots and primes mean derivatives with respect to $t$ and $y$
respectively. On the brane $y=0$ we assume there is a constant
tension $\sigma$, and an isotropic fluid with time-dependent
energy density $\rho=\rho(t)$. The energy-momentum tensor is now
\begin{equation}
T_{AB}=\delta(y)(-\sigma \delta_{\mu\nu}+\rho U_\mu U_\nu
+\tilde{p}h_{\mu\nu})\delta_A^\mu \delta_B^\nu. \label{35}
\end{equation}
Applying the junction conditions across the brane we obtain, for
arbitrary $y$, after integration with respect to $y$
\cite{brevik08a},
\begin{equation}
\left(\frac{\dot {a}}{na}\right)^2= \frac{1}{6}\Lambda+
\left(\frac{a'}{a}\right)^2+\frac{C}{a^4}, \label{36}
\end{equation}
\begin{equation}
H_0^2=\frac{1}{6}\Lambda+\frac{\kappa_5^4}{36}(\sigma+\rho)^2.
\label{37}
\end{equation}
We here let subscript zero refer to the brane. On the brane,
$n_0(t)=1$. Recall that $\Lambda$ and $\sigma$ are constants, and
that Eq.~(\ref{37}) is a 5D, not a 4D, equation. Its essential new
feature is that it contains a $\rho^2$ term. The equation
functions as a bridge between 4D and 5D cosmologies.

We observe the solution for $a_0(t)$ if $\rho=0$:
\begin{equation}
a_0(t)=e^{\sqrt{\lambda}\,t}, \quad
\lambda=\frac{1}{6}\Lambda+\frac{1}{36}\kappa_5^4 \,\sigma^2,
\label{38}
\end{equation}
normalized such that $a_0(0)=1$.

Inserting $\rho=\rho_*/X^{2\alpha}$ into Eq.~(\ref{37}) we get
\begin{equation}
H_0^2=\frac{1}{6}\Lambda+\frac{\kappa_5^4}{36}\left[
\sigma+\frac{\rho_*}{(1-BH_*t)^{2\alpha}}\right]^2. \label{39}
\end{equation}
Near the Big Rip, $t_s=1/(BH_*)$, the quantities $\Lambda$ and
$\sigma$ become unimportant, and we get
\begin{equation}
a_0(t) \sim \exp \left[\frac{(\kappa_5^2 /6)\rho_*}{(2\alpha-1)
(BH_*)^{2\alpha}(t_s-t)^{2\alpha-1}}\right], \label{40}
\end{equation}
showing that if $\alpha>1/2$, $a_0(t)$ has an essential
singularity. Einstein's gravity corresponds to $\alpha =1$. The
singularity becomes stronger, the higher is the value of $\alpha$.
If $\alpha < 1/2$, $a_0(t)$ does not diverge at $t_s$. From
Eq.~(\ref{36}),
\begin{equation}
a^2(t,y)=\frac{1}{2}a_0^2(t) \Bigg[ \left( 1+\frac{\kappa_5^4 \,
\sigma^2}{6\Lambda}\right) + \left(1-\frac{\kappa_5^4 \,
\sigma^2}{6\Lambda}\right) \cosh (2\mu\,y)-\frac{\kappa_5^2 \,
\sigma}{3\mu}\sinh (2\mu |y|) \Bigg], \label{41}
\end{equation}
with $\mu=\sqrt{-\Lambda/6}$. The important point here is that the
{\it Big Rip divergence on the brane becomes transferred to the
bulk}. The bulk scale factor $a(t,y)$ diverges for arbitrary $y$
at $t=t_s$ if $a_0(t)$ diverges at $t_s$. There is no fundamental
difference between an Einstein fluid and a modified gravity fluid
in this respect; their behavior is  essentially the same.

In summary, we have discussed viscous dark energy as a particular
representative of inhomogeneous equation-of-state fluids and  the
appearance of finite-time future singularities for such energies.
It is of interest to note that due to the relationship between
modified gravity and inhomogeneous equation-of-state ideal fluids
\cite{nojiri05a}, our findings may be useful in the study of
future singularities in modified gravity \cite{bamba08}.

%\newpage

\end{document}